\documentclass{iopart}
\usepackage{iopams}
\usepackage[sort&compress,numbers]{natbib}  
\eqnobysec

\newcommand{\prd}{Phys. Rev. D}
\newcommand{\pr}{Phys. Rev.}
\newcommand{\grg}{Gen. Rel. Grav.}
\newcommand{\apj}{Astrophys. J.}
\newcommand{\mnras}{Mon. Not. R. Astron. Soc.}
\newcommand{\prl}{Phys. Rev. Lett.}
\newcommand{\cqg}{Class. Quantum Grav.}

\newcommand{\prog}{Prog. Theor. Phys.}
\newcommand{\living}{Living Rev. Relativity}
\newcommand{\apjl}{Astrophys. J. Lett.}

\newcommand{\ajp}{Am. J. Phys.}
\newcommand{\annphys}{Annals Phys.}
\newcommand{\annalen}{Ann. Phys.}
\newcommand{\physzeit}{Phys. Zeit.}
\newcommand{\jmathphys}{J. Math. Phys.}
\newcommand{\amsci}{Am. Sci.}
\newcommand{\commathphys}{Comm. Math. Phys.}
\newcommand{\jhep}{J. High Energy Phys.}
\newcommand{\rmp}{Rev. Mod. Phys.}
\newcommand{\ijtp}{Int. J. Theor. Phys.}
\newcommand{\ijmpe}{Int. J. Mod. Phys. E}
\newcommand{\ssr}{Space Sci. Rev.}
\newcommand{\pasp}{Publ. Astron. Soc. Pac.}

\begin{document}

\review{The Kerr Metric}
\author{Saul A. Teukolsky}
\address{Center for Radiophysics and Space
    Research, Cornell University, Ithaca, New York 14853, USA}
\ead{saul@astro.cornell.edu}
\date{\today}

\begin{abstract}
This review describes the events leading up to the discovery of the Kerr metric
in 1963
and the enormous impact the discovery has had in the subsequent 50 years.
The review discusses the Penrose process,
the four laws of black hole mechanics, uniqueness of the solution, and the
no-hair theorems. It also includes Kerr perturbation theory and its
application to black hole stability and quasi-normal modes. The Kerr metric's
importance
in the astrophysics of quasars
and accreting stellar-mass black hole systems
is detailed. A theme
of the review is the ``miraculous'' nature of the
solution,
both in describing in a simple analytic formula the most general
rotating black hole, and in having unexpected mathematical properties
that make many calculations tractable. Also included is a pedagogical
derivation of the solution suitable for a first course in general
relativity.
\end{abstract}


\submitto{\CQG}
\maketitle

\section{Introduction}

The Schwarzschild solution was found within only
a few months of the publication of Einstein's field equations
\cite{schwarz1916}.
It is hard to imagine how different
the development of general relativity would have been without this exact
solution in hand. Instead of dealing only with small weak-field corrections
to Newtonian gravity, as Einstein had initially imagined would be the case,
fully nonlinear features of the theory could be studied, most notably
gravitational collapse and singularity formation.

The existence of the Schwarzschild solution set in motion a search for
other exact solutions. None was more eagerly sought than the metric
for a rotating axisymmetric source. Already in 1918,
Lense and Thirring \cite{lense1918}
had found the exterior field of a rotating sphere to first order in
the angular momentum, but was there a simple exact solution that was
physically relevant? It took almost 50 years to find such a solution:
the Kerr metric \cite{kerr1963}.
Now that another 50 years have elapsed, we can see what an enormous
impact this discovery has had. Practically every subfield of general
relativity has been influenced. And in astrophysics the discovery
of rotating black holes together with a simple way to treat their
properties has revolutionized the subject.

In this review, I will first give some forms of the metric for reference. Next
I will describe how Kerr found the solution, before moving on to
detailing its properties and applications.

\section{Forms of the Kerr metric}

Recall that a metric is \emph{stationary} if it has a Killing vector field
that is timelike at infinity. A metric is \emph{static} if it is
stationary and invariant under time reversal, or equivalently, if the
time Killing vector is hypersurface orthogonal. A rotating metric is
not invariant under $t\to -t$, and so must be stationary without being
static.

In Kerr's original paper, he presented the metric in the following
form\footnote{Kerr's coordinate $u$ is here denoted $v$ to be consistent
with the convention that $u$ denotes a retarded time ($u=t-r$ in flat
space) whereas $v$ denotes an advanced time ($v=t+r$ in flat space).
The sign of $a$ has also been corrected from the original paper, as Kerr
himself quickly noted later.}
\begin{equation}
\eqalign{
\rmd s^2 = -\left( 1 - \frac{2mr}{r^2+a^2\cos^2\theta}\right)\, \big(\rmd v
 - a \sin^2\theta \, \rmd \tilde\phi\big)^2\cr
\qquad{}+2 \big(\rmd v - a \sin^2\theta \, \rmd \tilde\phi\big) \,
\big(\rmd r - a \sin^2\theta \, \rmd \tilde \phi\big)\cr
\qquad{}+ \big(r^2+a^2\cos^2\theta\big)\, \big(\rmd\theta^2+\sin^2\theta\,
\rmd\tilde\phi^2\big).}
\label{eq:kerr1}
\end{equation}
Nowadays, it is easy to check with computer algebra
that this metric satisfies the vacuum
Einstein equations. Deriving it from some reasonable assumptions is
still not easy, however. I will return to the derivation in
\S\ref{sec:history} and \S\ref{sec:pedagog}.

When $a=0$, the above metric reduces to Schwarzschild in ingoing
Eddington-Finkelstein coordinates, so it is usually called the
ingoing Eddington-Finkelstein form of the Kerr metric.
The coordinate $\tilde\phi$ has a tilde to distinguish it from the
Boyer-Lindquist coordinate $\phi$ below. The ingoing principal null
vector (see \S\ref{sec:history})
is particularly simple in these coordinates---it is simply
$-\partial/\partial r$.
This form of the metric has three off-diagonal terms and so is quite
cumbersome for calculations.

The Schwarzschild curvature singularity at $r=0$ is replaced in the Kerr metric
by
$r^2+a^2\cos^2\theta=0$, that is, $r=0$ and $\theta=\pi/2$.
It is not exactly clear what the
geometry of this singularity is if we interpret $r$ and $\theta$ as
being like ordinary spherical polar coordinates.
The situation becomes clearer in the so-called Kerr-Schild
form, described next.

The Kerr-Schild form is very useful for finding exact solutions to the
field equations, although this is not how Kerr originally derived
the solution. A Kerr-Schild metric has the form
\begin{equation}
\rmd s^2=\left(\eta_{\alpha\beta} + 2H \ell_\alpha
\, \ell_\beta
\right)\rmd x^\alpha\, \rmd x^\beta
\end{equation}
where $\ell_\alpha$ is a null vector with respect to both
$g_{\alpha\beta}$ and $\eta_{\alpha\beta}$.
Kerr also gave the metric in this form in his original paper,
with
\begin{equation}
H=\frac{mr^3}{r^4+a^2 z^2},\qquad
\ell_\alpha =\left(1, \frac{rx+ay}{r^2+a^2}, \frac{ry-ax}{r^2+a^2},
  \frac{z}{r} \right).
\end{equation}
Here $r$ is not a coordinate but is implicitly defined by
\begin{equation}
\frac{x^2+y^2}{r^2+a^2} + \frac{z^2}{r^2} = 1.
\end{equation}
We now see that the singularity at
$r=0$ corresponds to the ring $x^2+y^2=a^2$, $z=0$.

A very convenient coordinate system for the Kerr metric was
introduced by
Boyer and Lindquist \cite{boyer1967} in 1967.
The transformation from ingoing Eddington-Finkelstein coordinates is
defined by
\begin{eqnarray}
\rmd v = \rmd t + (r^2+a^2)\,\rmd r/\Delta\\
\rmd \tilde\phi = \rmd \phi +a\,\rmd r/\Delta
\end{eqnarray}
where $\Delta\equiv r^2-2Mr+a^2$.
The metric in these coordinates has only one off-diagonal term:
\begin{equation}
\eqalign{
\rmd s^2 = - \left( 1- \frac{2Mr}{\Sigma}\right) \rmd t^2
- \frac{4Mar\sin^2\theta}{\Sigma}\, \rmd t\,\rmd\phi
+ \frac{\Sigma}{\Delta}\rmd r^2
+ \Sigma \,\rmd\theta^2\cr
\qquad{}+ \left(r^2+a^2+ \frac{2M a^2 r\sin^2\theta}{\Sigma}\right)
\sin^2\theta\,\rmd\phi^2}
\label{eq:bl}
\end{equation}
where $\Sigma\equiv r^2+a^2\cos^2\theta$.
In these coordinates, the metric is manifestly asymptotically flat, and
$M$ and $J=aM$ are easily identified as the mass and angular momentum
by letting $r\to\infty$.
When $a=0$, the metric reduces to Schwarzschild in standard curvature
coordinates.
Boyer-Lindquist coordinates will be used exclusively in the rest of
this paper.

Note that the Boyer-Lindquist form of the metric is singular also at
$\Delta=0$. This is a coordinate singularity, since the ingoing
Eddington-Finkelstein form is regular there. This behavior is reminiscent
of the situation in Schwarzschild, where $\Delta=0$ at $r=2M$, the
event horizon. In fact, it is easy to check that the
normal vector to surfaces $r={}$constant satisfies
\begin{equation}
n_\alpha n_\beta g^{\alpha\beta}=g^{rr}=\frac{\Delta}{\Sigma}.
\end{equation}
Accordingly, the normal vector is null when $\Delta=0$; the $\Delta=0$ surfaces
are null hypersurfaces. Examination of the null geodesics of Kerr
near these surfaces shows that they are in fact
horizons. The roots of $\Delta=0$ are
\begin{equation}
r_{\pm}=M\pm\sqrt{M^2-a^2}
\end{equation}
which defines the outer and inner horizons. As we will see later, the region
near $r_-$ is very likely not important physically, and so
I will refer to $r_+$ simply as the event horizon of the rotating black hole.

The charged generalization of the Kerr solution was found in 1965
\cite{newman1965} and is called the Kerr-Newman metric. In Boyer-Lindquist
coordinates, one simply replaces  $\Delta$ by $r^2-2Mr+a^2+Q^2$ to get
the solution.

\section{History of the discovery}
\label{sec:history}

Why was its so difficult to find a rotating generalization of the
Schwarzschild metric? The straightforward approach was
to follow Schwarzschild: Write down the most
general line element that reflected
the symmetries of the problem (stationarity and axisymmetry), then
get the field equations and try to solve them. But here Einstein's
instinct was correct: the equations were so complicated that nobody
succeeded. Lewis \cite{lewis1932} carried out the first notable attempt
in 1932.
Weyl in 1917 \cite{weyl1917} had simplified the static (nonrotating)
case by introducing ``canonical coordinates,'' and Lewis
showed that similar coordinates could be found in the rotating
case. The equations were simpler
in these coordinates, but still intractable in general. Lewis found
some solutions but, setting a pattern that persisted until Kerr's work,
none that was asymptotically flat and nonsingular.
A combination historical/personal account of this work
is given by Dautcourt \cite{dautcourt2009}. He singles out
several attempts to solve the Lewis or related equations
\cite{papapetrou1953,jordan1958,ehlers1962}, and describes his own futile
attempts.
As Ehlers and Kundt wrote in their
1962 review \cite{ehlers1962}, \emph{``the old problem of
constructing rigorously the field of a finite rotating body
is as yet unsolved, even as to its exterior part.''}
It was not until Ernst's reformulation
of the equations in 1968 \cite{ernst1968}
that the Kerr solution might possibly
have been discovered using this approach, had Kerr not already
done so five years earlier by a completely different method.

\subsection{Kerr's method of attack}
This section is based on an account given by Kerr himself
\cite{kerr2009} of the discovery
of the Kerr metric.

Kerr's success had its origin in a paper by Petrov in 1954 \cite{petrov1954}.
Petrov classified the algebraic properties of the Weyl tensor at
any point into three types plus some subcases. The classification
describes the properties of four null ``eigenvectors'' determined
by the Weyl tensor, now called principal null vectors.
If two or more of the eigenvectors coincide, the
metric is called \emph{algebraically special} in modern terminology.
Pirani \cite{pirani1957}
made researchers in the West aware of Petrov's work,
rechristening
Petrov's types and subtypes with distinct labels (Type I, Type D, \dots)
that have become standard. Kerr heard about this work
in a seminar by Pirani in 1957.

The Petrov classification had a huge impact on general relativity
theory, leading for example to the famous peeling theorem
for gravitational waves. An important result for the story
of the Kerr metric was the Goldberg-Sachs theorem \cite{goldberg1962}:
\emph{A vacuum metric is algebraically special if and only if it admits
a geodesic and shearfree null congruence.} Robinson and Trautman
\cite{robinson1962} used this result and the assumption that the congruence
was hypersurface orthogonal to reduce the complete solution of the field
equations to a single nonlinear PDE.

In 1962 Kerr met Alfred Schild at a meeting in Santa Barbara. Schild
invited Kerr to come to the newly founded Center for Relativity at the
University of Texas as a postdoc for the next academic year.

At this time,
Kerr was playing around with Einstein's equations using complex null
tetrads, a formalism mathematically equivalent to the Newman-Penrose
spinor formalism, which was introduced in that same year. Kerr
was trying to extend the Robinson and Trautman work to the case
where the null congruence of the algebraically special spacetime
was ``twisting,'' that is, not hypersurface orthogonal. The solutions
found by Robinson and Trautman were all static, and Kerr hoped that
generalizing the Robinson-Trautman approach to twisting congruences
might allow him to find the metric of a rotating source.

During this period, an important question being studied was the fate
of a star undergoing gravitational collapse. It was generally accepted
that a perfectly spherical star would collapse to a black hole described
by the Schwarzschild metric.\footnote{The term ``black hole'' was
actually only introduced by Wheeler in 1968 \cite{wheeler1968}.}
But was this merely an artifact of perfect symmetry? Maybe the slightest
angular momentum would halt the collapse before the formation of
an event horizon, or at least before the formation of a singularity.
Finding a metric for a rotating star would be very helpful in
answering these questions.

Kerr was not the only one pursing such a solution. Robinson and Trautman
presented some results on metrics with twisting null
congruences at the GR3 conference in Warsaw
in 1962 that Kerr attended, but still had not found any useful rotating
solutions \cite{robinson1964}. Then a preprint appeared by Newman,
Tabourino and Unti where they seemed to have proved
that the only algebraically
special spacetime with a twisting principal null congruence
is the so-called NUT space (after the authors' initials). This metric
is a one-parameter generalization of Schwarzschild that is not
asymptotically flat
except for the pure Schwarzschild case. If true, this would mean that
the rotating generalization of Schwarzschild would not be found among
the algebraically special solutions.
Kerr and his colleague Alan Thompson checked the Newman \etal paper and
found an error, which was corrected in the published version \cite{newman1963}.
So Newman \etal had not ruled out the possibility of algebraically special
rotating solutions; Kerr continued his search.

The earlier attempts to find a rotating axisymmetric solution by Lewis,
Papapetrou, and others, were based on using the symmetries to simplify
the metric and then trying to solve the resulting field equations. By
contrast, Kerr's method was to assume first that the metric would
be algebraically special like Schwarzschild, simplify the metric
accordingly (e.g., with the Goldberg-Sachs theorem),
and only then impose the $t$ and $\phi$ symmetries.
Although the simplification of the metric and field equations
for the algebraically special case now appears in standard
monographs (e.g., \cite{stephani2003}), it was not trivial
at the time. The next step, imposing the symmetries, was done
very cleverly. Instead of trying to solve Killing's equation directly,
which probably would have been impossible, Kerr used the action of the
symmetry group explicitly to deduce the allowed form of the
Killing vectors. Kerr imposed the $t$ and $\phi$ symmetries each in turn,
and found a solution that was asymptotically flat with two parameters,
$M$ and $a$. When $a=0$, the solution reduced to Schwarzschild.
Did a nonzero $a$ turn it into
the long-sought rotating solution?

Kerr describes how Schild sat excitedly in Kerr's office while Kerr
calculated the angular momentum of the solution. Schild puffed away
at his pipe while Kerr chain smoked cigarettes. Kerr transformed
the metric to coordinates in the standard form for asymptotically
flat metrics, \eref{eq:lense} below, read off the Lense-Thirring term,
and then announced to Schild
``It's rotating!'' The discovery paper \cite{kerr1963}
was sent to Physical Review
Letters in July 1963 and appeared in September.
It was only 1 1/2 pages long and gave little hint of how the
solution was found. 
To those not expert in the long history of failure
to find such a solution, the physical importance of the paper was
almost certainly not evident. And anyone trying to check the
solution in those pre-computer-algebra days would have been mystified.
In fact, the details only appeared seven years later \cite{kerr1970}.
An alternative derivation, starting from the Kerr-Schild
ansatz,
appeared earlier \cite{kerr1965}. Some of the details of the
rather terse Kerr-Schild paper are explained in \cite{krasinski2009}.

In December of that year (1963), Kerr went to the first Texas Symposium on
Relativistic Astrophysics in Dallas. The conference was prompted by
the discovery of quasars earlier in the year. Their high redshift
suggested that relativity might have something to do with their
underlying mechanism, so the conference was an attempt
to bring together
relativists and astrophysicists. Kerr's conference paper \cite{kerr1964}
was the first major announcement of the new metric. In the
paper, Kerr pointed
out that gravitational collapse to a Schwarzschild black hole
had difficulty in explaining the prodigious energy output of quasars
because of the ``frozen star'' behavior for distant observers. However,
the properties of the event horizon were different with rotation
taken into account, as shown with the newly-discovered Kerr solution.
(In modern language, a naked singularity was visible for $a>M$.)

Kip Thorne \cite{thorne2004} has given an amusing description
of how Kerr's presentation had absolutely zero impact
on the meeting's audience.
Papapetrou was so incensed that he stood up to lecture the
participants on the importance of Kerr's discovery, but without
effect. Ironically, the Kerr solution is generally accepted today
as underlying the explanation of quasars (see \S\ref{sec:bz}),
although not because it is the external metric of a rotating star nor
because it describes a naked singularity.

\section{Properties of the Kerr metric}
Many important properties of the Kerr metric were figured out remarkably
quickly, as I now describe.

\subsection{Geodesics: a surprise}
In a stationary axisymmetric metric, the existence of the two Killing vectors
$\partial_t$ and $\partial_\phi$ implies that the corresponding
momenta of a test particle, $p_t$ and $p_\phi$, are constants of the
motion. The Hamiltonian itself always gives a constant of the motion, but
for a complete solution of the geodesic equations in four dimensions
we require a fourth constant of the motion. Given the lack of any
obvious symmetry in $r$ and $\theta$ in the Kerr metric, there was
no reason to expect the geodesics to be completely integrable.
So it came as a complete surprise when in 1968 Carter \cite{carter1968}
showed that a fourth constant could be found because the
Hamilton-Jacobi equation for the geodesics was separable in $r$ and
$\theta$. 
The treatment of Kerr geodesics is now a standard feature of essentially
all relativity textbooks, so I will not discuss it further here.
Important for later developments, as we will see,
is that also in 1968 Carter noted the separability
of the scalar wave equation in the Kerr metric \cite{carter1968b}.

\subsection{Maximal extensions, matter sources, and Birkhoff's Theorem}
In 1965 Penrose
\cite{penrose1965} had published the first black hole singularity
theorem, showing that singularity formation was an inevitable result
of gravitational collapse, and not some special feature of spherical
symmetry where all the collapsing particles were aimed at the same point.
Penrose also introduced the idea of using conformally compactified
diagrams to describe the causal relations in a spacetime, and Penrose
diagrams were applied to black hole spacetimes by Carter \cite{carter1966}.
Analogous to the Kruskal diagram for Schwarzschild, the maximal
analytic extension of Kerr was worked out
\cite{boyer1967,carter1966,carter1968}, revealing a complicated
structure of ``universes'' patched together.

The modern viewpoint is that all the complications of the Kerr solution
beyond the inner horizon are irrelevant to real astrophysics. There
are two reasons for this. First,
the inner horizon is generically singular in the
sense of being an infinite blue-shift surface that magnifies perturbations
(see, e.g., \cite{penrose1968,poisson1989,poisson1990,droz1997}).
Second, there is no Birkhoff Theorem for the Kerr metric.
Outside a rotating star, the metric is \emph{not} described by Kerr.
As will be discussed below, a generic rotating star can have
gravitational multipoles that are not the same as Kerr.
The monopole and magnetic dipole moments of Kerr are described by
$M$ and $a$.
Kerr does have higher multipole moments, but
they are all expressible in terms of $M$ and $a$.
What we do expect is that if a rotating star collapses to a black hole,
then the exterior metric will asymptotically approach Kerr.
By contrast, during spherical gravitational collapse the exterior
metric is Schwarzschild at all times. This means that if we draw the
worldline of the surface of the collapsing star on the Kruskal diagram,
the region inside the horizon but outside the stellar surface is physical.
It does not make sense, however, to draw a similar diagram for Kerr.
The bottom line is that
talk of analytically extending metrics to expand out
into other universes has come to
be realized to be of mathematical interest at best.

A related issue is that of a possible interior matter solution for the
Kerr metric.
I will take the viewpoint that, while there may be some mathematical interest
in searching for such a solution, there is no physical interest. The reason
is again no Birkhoff's Theorem:
The Kerr metric does not represent the exterior metric
of a physically likely source, nor the metric during
any realistic gravitational collapse.
Rather, it gives the asymptotic metric at late times as whatever dynamical
process produced the black hole settles down.

\subsection{The Penrose process}
\label{sec:penroseproc}

In the Kerr metric, the time Killing vector $\partial_t$ changes from
being timelike to being spacelike in a region outside the event horizon.
The boundary of this region occurs where the Killing vector becomes null:
\begin{equation}
\partial_t \cdot \partial_t = g_{tt} = 0 \quad \Rightarrow \quad
r=M+\sqrt{M^2-a^2\cos^2\theta}.
\end{equation}
This surface is called the \emph{stationary limit} or the \emph{ergosurface}.
The region between the event horizon and the ergosurface is
the \emph{ergosphere}. The existence of the ergosphere allows various
kinds of energy extraction mechanisms for a rotating black hole.
For example, inside the ergosphere, a particle with
4-momentum $\mathbf{p}$ can have negative conserved energy $E=-\mathbf{p}
\cdot \partial_t$. In the Penrose process
\cite{penrose1969,penrose1971}
a particle falls into the ergosphere where it splits into two. One
of the particles is created with negative energy and falls into the
hole. The other comes out to infinity with more energy than the original
particle, with the black hole losing some rotational energy.

The original Penrose process is not likely to be astrophysically
important \cite{bardeen1972}. Its wave analogue, superradiant scattering,
is much more interesting (see \S\ref{sec:superradiant}).
And its manifestation in the Blandford-Znajek process for
electromagnetic energy extraction from black holes is at the foundation
of current explanations of prodigious astrophysical energy sources
such as quasars (see \S\ref{sec:bz}).

\section{The four laws of black hole mechanics}

\subsection{The classical era}
The development of black hole thermodynamics
began with analogies to the second law. Penrose and Floyd
\cite{penrose1971} had noted that the surface area of a Kerr black hole
increased during the Penrose process. Without giving details, they stated
``In fact, from general considerations one may infer that there should
be a natural tendency for the surface area of the event
horizon of a black hole to increase with time whenever the situation
is non-stationary.''
Independently, Christodoulou
\cite{christodoulou1970} had shown that a quantity he called
the irreducible mass could never decrease when particles were
captured by Kerr black holes, and investigated ``reversible transformations,''
where it remained unchanged. The decisive step came with Hawking's Area Theorem,
giving a proof of the Penrose-Floyd statement
that the surface area of a generic black hole could never decrease
\cite{hawking1971}. Bekenstein \cite{bekenstein1973},
who was also a graduate student
at Princeton and so very familiar with Christodoulou's thesis work,
proposed that black hole area was not just analogous to entropy,
but actually was proportional to its entropy. He based his argument
on information theory: The black hole entropy was a measure
of the information about the interior that was inaccessible to an
external observer. He argued that the
constant of proportionality should be the square of the Planck length, up to 
a numerical factor. Bekenstein also proposed a generalization of the second
law of thermodynamics: The sum of the black hole entropy plus the
ordinary entropy in the exterior never decreases.

At about the same time, the first law of black hole
mechanics was developed. For the Kerr metric, the area of the horizon is
\begin{equation}
A=8\pi M r_+.
\end{equation}
Replacing $a$ in this expression by
$J=aM$, the angular momentum of the hole, and differentiating gives
\begin{equation}
\delta M = \frac{1}{8\pi}\kappa \delta A + \Omega \delta J
\label{eq:firstlaw}
\end{equation}
where
\begin{equation}
\kappa=\frac{r_+-M}{2Mr_+},\qquad \Omega=\frac{a}{2Mr_+}.
\label{eq:angvel}
\end{equation}
Here $\Omega$ is the ``angular velocity of the horizon'' and $\kappa$
is the surface gravity, related to the acceleration of a particle
corotating with the black hole at the horizon. Bekenstein \cite{bekenstein1973}
and Smarr \cite{smarr1973} had noted the similarity between  
\eref{eq:firstlaw} and the first law of thermodynamics, where if $A$ was
proportional to the black hole entropy then $\kappa$ would have to be
proportional to the black hole temperature.

These ideas were synthesized and extended by Bardeen, Carter, and Hawking
\cite{bardeen1973b}. They proved the validity of the first law
\eref{eq:firstlaw} for general stationary black hole configurations,
including external matter distributions.
They also gave a proof of the zeroth law, that the surface gravity
$\kappa$ is constant on the black hole horizon. This gives further support
to the idea of $\kappa$ as related to temperature, which is constant
for a body in thermal equilibrium. They finally conjectured that the third law
would hold in the form that the temperature of a black hole, i.e., $\kappa$,
cannot be reduced to zero by a finite sequence of operations.
It follows from the third law that nonextremal black holes cannot be
made extremal in a finite number of steps.
The third law was later proved by Israel \cite{israel1986}.
Note that Planck's formulation of the third law of thermodynamics
does not hold in black hole mechanics: The entropy (area) of an
extremal black hole is finite even though it has zero
temperature (surface gravity).

By now, there exist many different proofs of these laws, making slightly
different assumptions (see \cite{wald2001} for a review
and references).

\subsection{The quantum era}
In the classical treatment, the laws of black hole mechanics were
considered to be only analogies to the real laws of thermodynamics.
For example, since a black hole could not radiate, it always had to
have zero temperature. This meant that $\kappa$ could not be the
actual temperature of the hole. All this changed with Hawking's discovery
that when quantum effects were considered, a black hole \emph{can}
radiate \cite{hawking1974,hawking1975}. Hawking showed that the radiation
was thermal with a temperature
\begin{equation}
T=\frac{\hbar\kappa}{2\pi}
\end{equation}
where Boltzmann's constant is set to unity in the denominator.
By the first law
\eref{eq:firstlaw}, this fixes the constant
of proportionality between area and entropy:
\begin{equation}
S=\frac{1}{4} \frac{A}{\hbar}.
\end{equation}
As Bekenstein had argued, the entropy was indeed proportional to the
area in Planck units. (Hawking had actually not been persuaded by Bekenstein's
original proposal, and had a different motivation for his calculation. See
\cite{page2005} for some historical remarks and references.)

The discovery of a rational basis for assigning entropy to black holes and
the formulation of
a generalized second law that includes black hole interactions
have spawned a huge amount of research that would take us too far afield to
discuss here. See the companion article by Jacobson \cite{jacobson2014}
and also \cite{grumiller2014} for reviews and references.
Perhaps the most
interesting question that has come out of this is what the entropy
of a black hole represents microscopically. If entropy is fundamentally
the log of the number of states accessible to the system, then what
are these accessible degrees of freedom for a black hole?
Some of the proposed ideas are discussed in \cite{page2005}, for example.
Perhaps the most intriguing comes from string theory,
where microstates counted for D-brane configurations allow the entropy
of certain black holes to be calculated.
Who could have imagined that, in such a short time,
Kerr's playing around with principal null vectors of the Weyl tensor
would lead
to direct explorations of quantum gravity in string theory?

\section{Black Hole Uniqueness}
In 1967 Werner Israel \cite{israel1967}
published a remarkable theorem: the only static vacuum
black holes are the Schwarzschild black holes (and hence are spherically
symmetric).\footnote{Here and in most of this review, I will gloss
over the exact assumptions that go into proving the various theorems
that I quote. Among other things, it is here assumed that one is
dealing with asymptotically flat black holes with no cosmological constant.}
Unlike equilibrium matter objects,
it was not possible to find a static black hole with
a quadrupole moment, for example. Israel quickly extended his result
to a uniqueness theorem for static black holes with charge: only
Reissner-Nordstrom black holes are allowed \cite{israel1968}.

Israel's paper set off a burst of activity to see if the uniqueness result holds
also for the stationary case: does the Kerr metric describe all
possible rotating black holes? This question is related to the
conjecture that a black hole formed by gravitational collapse will
asymptotically settle down to a member of the Kerr family.
(We usually assume that the charge of a real black hole is
negligible because macroscopic
astrophysical charged objects will
rapidly be neutralized by surrounding plasma. However, in theoretical
general relativity the uniqueness
of the Kerr-Newman family has also been an important question.)

More precisely, the conjecture was that the only stationary, asymptotically
flat solution of the vacuum Einstein equations that is nonsingular from
infinity to a regular event horizon is the Kerr metric.

Carter \cite{carter1971} showed that axisymmetric black holes could depend
on only two parameters, the mass and angular momentum. He also
showed that the Kerr metric
was the only one that included a zero angular momentum black hole.
(More precisely, he showed that there were no linear perturbations
near Kerr that were stationary and axisymmetric other than changes
in the mass and angular momentum.)
These results strongly suggested
that Kerr black holes were the only ones.
Next, Hawking \cite{hawking1972} proved that all stationary black holes
must be either static or axisymmetric, making one of Carter's
assumptions unnecessary. He also
showed that the topology of the event horizon had to
be spherical, another assumption that was used in the Israel/Carter
work. Finally, in a spectacular feat of ingenuity, Robinson \cite{robinson1975}
gave a definitive proof of Carter's results without comparing Kerr
only to nearby solutions.

For most physicists and astronomers, this was the end of the subject.
Kerr black holes were all they needed to understand. However, 
among the more mathematically inclined the fun had only begun.
Kerr-Newman uniqueness was finally nailed down in
the 1980's by Mazur \cite{mazur1982} and Bunting \cite{bunting1983}.
Their work could also be used to simplify and improve the earlier results
for uncharged black holes. 
Interestingly, in the case of extremal charged black holes, with
$Q^2=M^2$, the uniqueness theorem also allows Majumdar-Papapetrou
solutions \cite{hartle1972}.
These solutions describe an arbitrary number of black
holes in equilibrium with the gravitational attraction exactly balanced
by Coulomb repulsion.

Much of the work since the 1970's has focused on removing the
``technical'' assumptions made in the earlier proofs and
on improving the mathematical rigor, but there is still
the sense that some of the assumptions are not necessary. Excellent
reviews are given in
\cite{chrusciel2012,heusler1996,carter1999,robinson2009}.
The main fly in the ointment today is still the cosmic censorship
hypothesis: if singularities can occur outside an event horizon, then
the proofs fail.
The hypothesis was introduced by Penrose \cite{penrose1969},
who
postulated the existence of a ``cosmic censor'' who
forbids the appearance of ``naked'' singularities not clothed by
an event horizon.
Despite the failure to find any convincing example
of a naked singularity forming from well-behaved generic initial data,
there is still no proof that this is impossible.

The uniqueness theorems have led to two strands of ongoing research:
generalizations of the no-hair theorem, and questions of uniqueness
for black holes in higher dimensions.

\subsection{No-hair theorems}

In parallel with the work on black hole uniqueness in the 1960's,
the community's understanding of gravitational collapse
began to  shift away from the frozen-star viewpoint.
With a comoving vantage point instead, the picture that developed was that
the gravitational field decouples from its matter source in the
late stages of collapse and radiates away all the multipoles it can,
leaving only the ``charges'' associated with a conserved flux integral
at infinity.

This idea was catalyzed by Doroshkevich,
Zel'dovich and Novikov \cite{dzn1966}, who showed that
the higher multipole moments of a system died out during gravitational
collapse---the beginnings of the no-hair theorem. Exactly how this
occurred only became clear later, with Price's
Theorem \cite{price1972a,price1972b}:
``Whatever can be radiated is radiated.'' Price gave quantitative
estimates of the decay of multipole moments measured at infinity.

By 1969 Penrose could write \cite{penrose1969}:
\begin{quote}
Doubts have frequently been expressed concerning
[the no-hair conjecture], since it
is felt that a body would be unlikely to throw off all its excess
multipole moments just as it crosses the Schwarzschild radius.
\dots On the other hand, the gravitational field \emph{itself} has
a lot of settling-down to do after the body has fallen into the
``hole''. The asymptotic measurement of the multipole moments need have
very little to do with the detailed structure of the body itself; the
\emph{field} can contribute very significantly. In the process of settling
down, the field radiates gravitationally---and electromagnetically
too, if electromagnetic field is present. Only the mass, angular
momentum and charge need survive as ultimate independent parameters.
\end{quote}

Wheeler introduced the aphorism  ``black
holes have no hair'' in 1971 \cite{wheeler1971}.

More recent work has made it clear that the matter fields
play a crucial role in the applicability of the no-hair theorem.
A trivial example is that one can construct stationary black holes
surrounded by equilibrium rotating
disks of perfect fluid or collisionless matter.
However, we now know that
even matter fields analogous to electromagnetic fields, with
generalized conserved charges, can violate the no-hair theorem.
In particular, static solutions for
Yang-Mills fields coupled to gravity can not be classified simply by
mass, angular momentum, and conserved charges. In fact, a wide variety of
scalar field models violate the no-hair theorem. Since these
scalar fields are precisely of the type that are expected to
occur in the low-energy limit of string theory, they are not
just of mathematical interest. The situation is reviewed
in \cite{chrusciel2012}.
These violations of the no-hair theorem are static and spherically
symmetric. There is also numerical evidence
that with suitable matter fields, static black holes are not necessarily
spherically symmetric or even axisymmetric, and
nonrotating black holes need not be static.
Whether these results are astrophysically important or simply points
of principle is not clear.

Interestingly, experimental tests of the no-hair theorem have been
proposed. The idea is to measure the spin and quadrupole moment of
a black hole and see if they satisfy the Kerr relationship. See,
for example, \cite{will2008,sadeghian2011}.

\subsection{Black holes in higher dimensions}
Interest in black holes in higher dimensions is motivated by string
theory, which describes gravity and requires more than four dimensions.
But this is not the only motivation: We learn about black holes and
general relativity in four dimensions by studying what happens in other
dimensions. In four dimensions, black hole properties include
\begin{itemize}
\item
uniqueness
\item
no-hair theorem, i.e., characterization by conserved charges
\item
spherical topology
\item
dynamical stability
\end{itemize}
\emph{None} of these properties holds in higher dimensions.
Excellent reviews of these and other properties are given in
\cite{emparan2008,horowitz2012,hollands2012}.

An example of a regular, stationary,
asymptotically-flat, vacuum solution in five dimensions
is the Myers-Perry black hole \cite{myers1986}. Such solutions in
fact exist in all higher dimensions and are
generalizations of the Kerr metric. They have spherical horizon
topology and can be rotating in several independent
rotation planes.  But in five dimensions there are also
the Emparan-Reall black rings
with $S^2 \times S^1$ horizon topology \cite{emparan2002};
the Pomeransky-Senkov
black rings, which have a second
angular momentum parameter \cite{pomeransky2006}; and the ``Black Saturn''
solutions, which are
multi-component black holes where a spherical horizon
is surrounded by a black ring \cite{elvang2007}.

Looking at these solutions, we see that
there are black rings and Myers-Perry
black holes with the same mass and angular momentum.
In fact, black rings are not fully characterized by their conserved charges.
The
Black Saturns are regular vacuum multi-black-hole
solutions, which do not to exist in four dimensions.
Clearly the situation in higher dimensions is much more complicated
than in four dimensions.

\subsection{Numerical relativity and black hole uniqueness}

In the past ten years, fully 3-d numerical computations with black
holes have become possible. A natural question is to ask whether
the final state of, for example, the inspiral and merger of a binary
black hole system is a Kerr black hole. The question has
been studied most thoroughly by Owen \cite{owen2009,owen2010}, who
gives references to earlier work.

Owen \cite{owen2009} used numerical data from a high-precision
simulation of an equal-mass, nonspinning black hole-black hole binary
carried out with the \texttt{SpEC} code
\cite{scheel2009}. He devised a (spatially) gauge invariant definition
of multipoles for nonaxisymmetric dynamical metrics, and confirmed that the
multipoles settle down to the expected Kerr values. The agreement
could be measured to
at least a part in $10^5$ because of the high accuracy of the numerical
code. Owen also showed that that the exponential falloff of
the dominant quasi-normal mode at late times agreed with the expected
Kerr value. (Quasi-normal modes are discussed in \S\ref{sec:qnm}.)
Higher quasi-normal modes are more difficult to analyze because
the assignment of $l$ mode numbers
to the modes is not straightforward. (See \S\ref{sec:perts} for a discussion
of angular modes of Kerr.)
Accordingly, Owen also analyzed the quasi-normal modes of a head-on collision
between two black holes, one with spin up and one with spin down
\cite{lovelace2010}. Here the final state is expected to be
Schwarzschild, which makes the assignment of $l$ mode numbers
to the modes unambiguous.
Again he found agreement limited only by the numerical accuracy of the
simulations.

In \cite{owen2010}, Owen extended work by Campanelli \etal
\cite{campanelli2009} to confirm that the spacetime of
a black hole merger approaches Petrov Type D at late times.

\section{Pedagogical Derivation of the Kerr Metric}
\label{sec:pedagog}
As we have seen, the original derivation of the Kerr metric was
a tour de force of complicated calculations. 
Even the alternative derivation in the Kerr-Schild paper \cite{kerr1965}
admitted,
``It is well worth pointing out that the calculations giving these results
are by no means simple.'' This causes a problem for teaching a course
in the subject. One can simply present the metric to students without
derivation, or have them verify with computer algebra that it satisfies
the vacuum Einstein equations. However, it would be nice if there
were a pedagogical derivation on a par with the derivation of the
Schwarzschild solution that is usually done in a course. By ``pedagogical''
I mean that computer
algebra is allowed for computing the Einstein or Ricci tensor
from a given metric (e.g., using the
Mathematica program at the website \cite{hartle2003}), but that the assumptions
and techniques have to be suitable for beginning students.
So, for example, the derivation in
Chandrasekhar's book \cite{chandra1983} is ``straightforward'' in the sense
that it does not make use of concepts
like algebraic specialness. But even with evaluation of the field equations via
computer algebra, the subsequent solution is far from trivial.

There have been several attempts in the literature to formulate a
pedagogical derivation.
Enderlein \cite{enderlein1997}
makes use of the Lorentz-transformed basis of 1-forms
for a flat space—time in oblate spheroidal coordinates.
However, the field equations still require complicated manipulations to solve.
Deser and Franklin \cite{deser2010} start with a rather obscure
(for beginning students) motivation of the metric. They then use
``Weyl's trick'' to simplify the variational principle for
the field equations. (This trick is to impose the symmetries before
carrying out the variation, which under certain conditions gives the
correct final equations.) It is necessary to manipulate the scalar curvature
$R$ in the integrand of the variational principle by suitable integrations
by parts to make the problem tractable.
While suitable for experts, this does not meet my definition of pedagogical.

Dadhich \cite{dadhich2013} does not actually use the field equations.
Instead, he makes some
arbitrary-seeming assumptions about geodesics. However, he
starts with a good form for the metric.

It seems that the only way to construct a derivation transparent enough for
a first course is to make some heuristic assumptions that allow one to
start with a metric in suitable form. The field equations that follow from
the assumed metric must be simple enough to be solved and yield the desired
metric. The question of what heuristic assumptions are plausible is, of
course, subjective and also tainted by knowing the desired answer.
Here I will assemble what I consider to be the best elements of earlier
attempts into what I would use to present a derivation to beginning
students.

The metric outside a weak-field, slowly rotating source is approximately
\begin{equation}
\fl
\rmd s^2=-\left(1-\frac{2M}{r}\right)\rmd t^2 +\left(1+\frac{2M}{r}\right)
\rmd r^2 + r^2\rmd\theta^2+
r^2\sin^2\theta\,\rmd\phi^2
-\frac{4aM}{r}\sin^2\theta\,\rmd\phi\,\rmd t
\label{eq:lense}
\end{equation}
to first order in $a$.
Here $M$ is the mass and $a=J/M$ is the angular momentum per unit mass.
The last term describes the Lense-Thirring ``dragging of inertial frames.''
Grouping the mass terms in $g_{tt}$ and $g_{t\phi}$
together suggests rewriting this as
\begin{equation}
\fl
\rmd s^2=-\rmd t^2 +\left(1+\frac{2M}{r}\right)\rmd r^2 + r^2\rmd\theta^2+
r^2\sin^2\theta\,\rmd\phi^2+
\frac{2M}{r}(\rmd t-a\sin^2\theta\,\rmd \phi)^2,
\label{eq:lense2}
\end{equation}
but still only really valid to first order in $a$.

Now consider the effect of rotation on the spatial geometry.
With no rotation, the Schwarzschild geometry is spherically symmetric.
We expect rotation to ``flatten'' the geometry and so we want to choose
coordinates in which this effect will appear simple.
Following \cite{krasinski1978}, the simplest choice
is to use ellipsoids to define the coordinate system. In axisymmetry,
a family of confocal ellipsoids $r={}$constant is defined by
\begin{equation}
\eqalign{
x=\sqrt{r^2+a^2}\sin\theta\cos\phi \cr
y=\sqrt{r^2+a^2}\sin\theta\sin\phi \cr
z=r\cos\theta.}
\end{equation}
The flat space  metric in these oblate spheroidal coordinates is
\begin{eqnarray}
\fl
\label{eq:ellipsoidal}
-\rmd t^2+\rmd x^2+\rmd y^2 +\rmd z^2=-\rmd t^2+\frac{\Sigma}{r^2+a^2}\rmd r^2
+\Sigma\,\rmd \theta^2+(r^2+a^2)\sin^2\theta\,\rmd \phi^2\\
\fl
=-\frac{r^2+a^2}{\Sigma}(\rmd t-a\sin^2\theta\,\rmd \phi)^2
+\frac{\Sigma}{r^2+a^2}\rmd r^2
+\Sigma\,\rmd \theta^2
+\frac{\sin^2\theta}{\Sigma}[(r^2+a^2)\rmd\phi-a\,\rmd t]^2
\label{eq:ellipsoidal2}
\end{eqnarray}
where $\Sigma=r^2+a^2\cos^2\theta$.
The form \eref{eq:ellipsoidal2} is a simple rewriting of \eref{eq:ellipsoidal}.

Although the $\rmd t\,\rmd\phi$ terms actually
cancel out of \eref{eq:ellipsoidal2},
it provides a good starting point for finding a rotating metric, where
these terms do not cancel.
Comparing \eref{eq:lense2} and \eref{eq:ellipsoidal2} suggests making
the ansatz of replacing $r^2+a^2$ in the first and second terms
of \eref{eq:ellipsoidal2} with arbitrary functions of $r$. We could make
the more general ansatz of arbitrary functions of $r$ and $\theta$, but
that turns out to be too complicated. So we try the simpler assumption first.
In particular, let the coefficient of the first term in \eref{eq:ellipsoidal2}
become
\begin{equation}
\frac{r^2+a^2}{\Sigma}\to\frac{Y(r)}{\Sigma}\equiv\frac{r^2+a^2-Z(r)}{\Sigma}
\end{equation}
while in the second term let $r^2+a^2\to F(r)$. Then undoing the
operation\footnote{It turns out that using the form \eref{eq:ellipsoidal}
directly with the replacements $Y(r)$ and $F(r)$ is also tractable and only
slightly more complicated than using $Z(r)$ and $F(r)$.}
that led from \eref{eq:ellipsoidal} to \eref{eq:ellipsoidal2} gives
\begin{equation}
\fl
\rmd s^2=-\rmd t^2+\frac{\Sigma}{F(r)}\rmd r^2
+\Sigma\,\rmd \theta^2+(r^2+a^2)\sin^2\theta\,\rmd \phi^2
+\frac{Z(r)}{\Sigma}(\rmd t-a\sin^2\theta\,\rmd \phi)^2.
\label{eq:ansatz}
\end{equation}
Note that in getting from \eref{eq:lense2} to \eref{eq:ansatz}
I have made the unjustified assumption that the
quantity $a$ in the Lense-Thirring term is the same as the parameter
$a$ describing the oblateness of the spheroidal coordinates.

Now compute the vacuum field equations $G_{\mu\nu}=0$
from the metric \eref{eq:ansatz}
using a computer algebra program (e.g., the Mathematica program
\cite{hartle2003}).
To avoid complications with trigonometric functions, let $q=\cos^2\theta$
so that $\rmd\theta^2=\rmd q^2/[4q(1-q)]$.
All the equations are quite complicated, except for the equation
$G_{rr}=0$. This equation has a coefficient of $q$ and  a constant term,
which must separately vanish. This gives two equations:
\begin{eqnarray}
\label{eq:eins1}
-r^2(r^2+a^2-Z)+[Z+r(r-Z')]F=0\\
(r^2-Z)(r^2+a^2-Z)+r(Z'-r)F=0.
\label{eq:eins2}
\end{eqnarray}
These equations are equivalent to those solved in \cite{deser2010}.
Solving \eref{eq:eins2} for $F$ gives
\begin{equation}
F=\frac{(r^2-Z)(r^2+a^2-Z)}{r(r-Z')}.
\label{eq:F}
\end{equation}
Substituting for $F$ in \eref{eq:eins1} and ignoring the spurious
solution $Z=r^2+a^2$ (which would give a singular metric with $F=0$),
we find
\begin{equation}
Z'=Z/r \qquad \Rightarrow\qquad Z=2Mr, \qquad F=r^2-2Mr+a^2.
\label{eq:soln}
\end{equation}
The constant of integration in solving for $Z$ is set by comparing
with the weak-field limit \eref{eq:lense2} when $r\to\infty$.
One can easily check that the remaining components of
the Einstein tensor are zero for the solution \eref{eq:soln}.
The final form of the solution is the Boyer-Lindquist metric
\eref{eq:bl}.

\section{Kerr Perturbations}
\label{sec:perts}

The study of black hole perturbations was initiated
in a groundbreaking paper by Regge and Wheeler
in 1957 \cite{regge1957}.
Their goal was to prove that the Schwarzschild metric described a solution
that was stable to linear perturbations. The spherical symmetry and
time independence of the metric allowed the perturbations to be decomposed
into spherical harmonics and Fourier modes, and they derived a radial
equation describing the odd-parity perturbations. This work was
brought to fruition
by Vishveshwara \cite{vishveshwara1970} and Zerilli
\cite{zerilli1970,zerilli1970b}
in the early 1970's. In particular, Zerilli derived a radial equation
for the even-parity modes that allowed the stability proof to be completed.
The theory of Schwarzschild perturbations has turned out to be rich
in unexpected applied mathematics as well as in practical applications.
Much of this work in summarized in Chandrasekhar's book \cite{chandra1983}.

With the success of Vishveshwara and Zerilli for the Schwarzschild case,
attention naturally turned to Kerr.
While it was clear that one would still be able to separate out the
$t$ and $\phi$ dependence in the perturbation equations, there was no
obvious reason why the $r$ and $\theta$ dependence would separate.
Nevertheless, Carter's discovery \cite{carter1968b}
of the unexpected separability of
the scalar wave equation on the Kerr background gave some hope.

Rather than tackle the full gravitational perturbation problem
head on, Fackerell and Ipser \cite{fackerell1972} decided to look first
at Maxwell's equations on a Kerr background. They used the Newman-Penrose
(NP) formalism, inspired by Price's work on Schwarzschild perturbations
\cite{price1972b}. Price had used the NP formalism to recover the
Regge-Wheeler equation in a relatively simple way and to prove
important results about the late-time decay of perturbations.
Since the explicit form of the equations in NP language can be
chosen to encode the principal null vectors, it is reasonable
to expect the formalism to be well-suited to Type D metrics like
Schwarzschild and Kerr.
In the NP formalism, the gravitational field is described by
an overdetermined set of equations that includes
the Bianchi identities for the Weyl tensor explicitly. The ten independent
components of this tensor appear as five complex quantities,
$\Psi_0$, $\Psi_1$, $\Psi_2$, $\Psi_3$, $\Psi_4$.
In Price's work, the Regge-Wheeler equation appeared as
the equation governing the imaginary part of the ``middle'' quantity,
$\Psi_2$.

For Maxwell's equations, the NP formalism uses
three complex quantities $\Phi_0$,
$\Phi_1$, and $\Phi_2$, which encode the six components of $\mathbf{E}$ and
$\mathbf{B}$.
Fackerell and Ipser combined these equations on the Kerr background
and found a decoupled equation for the middle quantity $\Phi_1$.
Unfortunately, this equation did not separate in $r$ and $\theta$ and
so was of limited use. The gravitational problem still
appeared out of reach.

Meanwhile, Bardeen and Press \cite{bardeen1973}
had been collaborating on 
using Price's NP approach to study
radiation fields on the Schwarzschild background.
At this time, I was
a beginning second-year graduate student at Caltech and not
making much progress on the problem that my advisor, Kip Thorne,
had given me.
One day, Press, who was a year ahead of me,
told me that Bardeen had found decoupled equations in Schwarzschild
for the ``extreme'' quantities $\Psi_0$ and $\Psi_4$ (and $\Phi_0$ and $\Phi_2$
in the electromagnetic case). Since Price, Fackerell and Ipser had all
been at Caltech, I was very familiar with their work and I wondered if
there might be decoupled equations for
the extreme quantities in Kerr as well.
In the NP formalism there are certain commutation relations
that are used to
effect the decoupling, and I suspected that the decoupling might
go through in any Type D metric if the choice of null vectors
was all that mattered. 
The calculation took only a few hours, and indeed there were decoupled
equations. I was quite excited at first, but when I wrote out the equation
for $\Psi_4$ in Boyer-Lindquist coordinates it did not separate,
just like the Fackerell-Ipser equation for $\Phi_1$.

For the next few weeks I tried all sorts of algebraic contortions to
see if I could make the equation separate. There didn't seem to be any
useful information in the mathematical literature on separability
of PDEs beyond the usual Killing vector-based separability and the
classification of separable coordinate systems for Poisson's equation.
Eventually Kip Thorne decided to send me to Maryland for a month to
consult with Charles Misner, an expert in PDEs (among many other things). I
still made no progress, and returned feeling that I had somehow let
Kip down.
I put the work aside and took up other problems
in an effort to finish my thesis. Every few weeks I would take out my
notes and try a few more manipulations. One evening, about six months
after finding the decoupling, something made me try a new set of
substitutions. A key variable in the NP formalism is $\rho$, whose
real and imaginary parts encode the divergence and curl of the
outgoing principal null congruence. In the Kerr metric in Boyer-Lindquist
coordinates,
\begin{equation}
\rho=-\frac{1}{r-ia\cos\theta}.
\end{equation}
I found that if I considered the equation satisfied
by $\rho^{-4}\Psi_4$ instead of $\Psi_4$, the equation suddenly separated.
Similarly, the electromagnetic equation for $\rho^{-2}\Phi_2$ was separable.
Interestingly, the $\Psi_0$ and $\Phi_0$ equations were separable without
modification. However, I had never looked at them before because
$\Psi_4$ and $\Phi_2$ describe outgoing as opposed
to incoming radiation at large $r$
and so were considered to be physically more relevant.
Had I looked at $\Psi_0$ or $\Phi_0$ earlier, I could have saved myself
a lot of anguish.

The final result could be written as a single master equation describing
all the perturbations \cite{teukolsky1972,teukolsky1973}:
\begin{eqnarray}
\fl \left[\frac{(r^2+a^2)^2}{\Delta}-a^2\sin^2\theta\right]\frac{\partial^2
\psi}{\partial t^2} +\frac{4Mar}{\Delta}\frac{\partial^2\psi}{\partial t
\partial\phi}+\left[\frac{a^2}{\Delta}-\frac{1}{\sin^2\theta}\right]
\frac{\partial^2\psi}{\partial\phi^2}\nonumber\\
\fl\qquad
 {}-\Delta^{-s}\frac{\partial}{\partial r}\left(\Delta^{s+1}\frac{\partial\psi}
{\partial r}\right)
-\frac{1}{\sin\theta}\frac{\partial}{\partial \theta}\left(\sin\theta
\frac{\partial\psi}{\partial\theta}\right)
-2s\left[\frac{a(r-M)}{\Delta}+\frac{i\cos\theta}{\sin^2\theta}\right]
\frac{\partial\psi}{\partial\phi}\nonumber\\
\fl\qquad
 {}-2s\left[\frac{M(r^2-a^2)}{\Delta}-r-ia\cos\theta\right]
\frac{\partial\psi}{\partial t}
+(s^2\cot^2\theta-s)\psi=4\pi\Sigma T.
\label{eq:master}
\end{eqnarray}
Here the spin weight parameter $s$ takes on the values 0, $\frac{1}{2}$, 1,
and 2 for scalar, neutrino, electromagnetic, and gravitational perturbations.
The quantity $\psi$ is the field (scalar, $\rho^{-4}\Psi_4$, etc.), and
the explicit form of the source terms $T$ is given in \cite{teukolsky1973}.
(The neutrino equation was given independently by Unruh \cite{unruh1973}.)

Consider first the vacuum case ($T=0$).
Inspection of \eref{eq:master} shows that it will separate in the form
\begin{equation}
\psi=e^{-i\omega t}e^{im\phi}S(\theta)R(r).
\end{equation}
The equations for $S$ and $R$ are
\begin{eqnarray}
\label{eq:angular}
\fl\frac{1}{\sin\theta}\frac{\rmd}{\rmd \theta}\left(\sin\theta
\frac{\rmd S}{\rmd\theta}\right)
+\left(a^2\omega^2\cos^2\theta-2a\omega s\cos\theta-\frac{(m+s\cos\theta)^2}
{\sin^2\theta}+s+A\right)S=0\\
\fl\Delta^{-s}\frac{\rmd}{\rmd r}\left(\Delta^{s+1}\frac{\rmd R}
{\rmd r}\right)
+\left(\frac{K^2-2is(r-M)K}{\Delta}+4is\omega r -\lambda\right)R=0
\label{eq:radial}
\end{eqnarray}
where $K\equiv (r^2+a^2)\omega-am$ and $\lambda\equiv A+a^2\omega^2-2am\omega$.
The angular equation \eref{eq:angular} is an eigenvalue equation for
the separation constant $A$. The eigenfunctions ${}_s S_{lm}(\theta)$ are
called spin-weighted spheroidal harmonics, since they generalize
spheroidal wave functions (when $s=0$) and spin-weighted spherical harmonics
(when $a\omega=0$). The radial equation \eref{eq:radial} is in effective
potential form, except the potential is complex and not short-range, unlike
the Regge-Wheeler or Zerilli equations.

When sources are present ($T\neq 0$), one can use the angular eigenfunctions
to expand the right-hand side of \eref{eq:master}. This gives a radial
equation identical to \eref{eq:radial} but with a source term on the right-hand
side. Boundary conditions and the computation of energy fluxes for the
radial equation are discussed in \cite{teukolsky1973,teukolsky1974}.
In particular, for outgoing waves at infinity, $\Psi_4$ is related
to the transverse traceless metric perturbation by
\begin{equation}
\Psi_4=-\case{1}{2}\omega^2
(h_{\hat \theta\hat\theta}-ih_{\hat\theta\hat\phi})
\label{eq:psih}
\end{equation}
and encodes the two polarization states of a gravitational wave.

The Dirac equation turned out to be tricky. However, Chandrasekhar
\cite{chandra1976} showed how to handle it by first separating
the variables and then decoupling the field components.

In the remainder of this section I will describe some applications of
the Kerr perturbation equations and some later developments of the
theory.

\subsection{The miraculous identities}
\label{subsec:teukstar}

Suppose that one has solved the equation for a gravitational
perturbation $\Psi_4$ (or an electromagnetic perturbation
$\Phi_2$). Then the corresponding quantity $\Psi_0$ (or $\Phi_0$)
should be completely determined without having to re-solve the
master perturbation equation, and vice versa. It is in fact possible to relate
the quantities directly by means of some remarkable identities
that follow from Einstein's or Maxwell's equations \cite{teukolsky1974}.

Let us first display the identities for the electromagnetic case.
Following the normalization conventions
of Chandrasekhar \cite{chandra1983},
write the separated quantities in the form
\begin{equation}
\Phi_0=R_1(r)S_1(\theta), \qquad
\case{1}{2}\rho^{-2}\Phi_2= R_{-1}(r)S_{-1}(\theta)
\end{equation}
where $S$ and $R$ satisfy \eref{eq:angular} and \eref{eq:radial} and
$S$ is normalized to unity over the sphere. The subscripts $\pm 1$
denote the spin weight $s$ of the quantities.
Now define the radial and angular operators
\begin{equation}
\mathcal{D}_n=\partial_r-\frac{iK}{\Delta}+2n\frac{r-M}{\Delta},
\qquad \mathcal{L}_n=\partial_\theta+Q+n\cot\theta
\end{equation}
where $Q=-a\omega\sin\theta+m/\sin\theta$ and $K$ defined
after \eref{eq:radial} is the negative of
the variable defined by Chandrasekhar. The ``adjoints'' of these
operators are defined by changing the signs of $K$ and $Q$ and
are denoted by $\mathcal{D}^\dagger$ and $\mathcal{L}^\dagger$.
Then Maxwell's equations imply
\begin{eqnarray}
\Delta\mathcal{D}_0\mathcal{D}_0 R_{-1}=\mathcal{C}\Delta R_{1}\qquad &
\Delta\mathcal{D}^\dagger_0\mathcal{D}^\dagger_0 \Delta R_{1}=
\mathcal{C} R_{-1}\qquad \\
\mathcal{L}_0\mathcal{L}_1S_1=\mathcal{C}S_{-1}&
\mathcal{L}^\dagger_0\mathcal{L}^\dagger_1S_{-1}=\mathcal{C}S_{1}
\end{eqnarray}
where the constant $\mathcal{C}$ is given by
\begin{equation}
\mathcal{C} = \sqrt{\lambda_{\rm Ch}^2+4ma\omega-4a^2\omega^2}.
\label{eq:starconst}
\end{equation}
Here $\lambda_{\rm Ch}=\lambda+s^2+s$ is the separation constant
used by Chandrasekhar, which has the advantage that it is independent
of $s$. We see that the operators $\mathcal{D}_n$ and $\mathcal{L}_n$
and their adjoints act like raising and lowering operators for
the radial and angular eigenfunctions.

Similar relations follow from the perturbed Einstein equations for
the $s=\pm 2$ functions \cite{teukolsky1974,chandra1983},
but now requiring four operators to change
$s=2$ to $s=-2$, for example.
Chandrasekhar \cite{chandra1983} has called these various
identities the Teukolsky-Starobinsky identities.
For $s=\pm 2$,
a new constant analogous to \eref{eq:starconst} appears, which turns out to
be complex. Chandrasekhar has called constants like \eref{eq:starconst}
Starobinsky constants.\footnote{These constants first appeared, without
derivation, in
a paper by Starobinsky and Churilov \cite{starobinsky1974}. A reader
consulting this paper may be confused to see that only the electromagnetic
constant is given. The much more complicated gravitational constant
is mentioned, but not explicitly given. The resolution of this puzzle
is that in the pre-Internet era, preprints of papers were mailed around,
and the preprint contained the gravitational
expression. However, the expression was inexplicably omitted
in the published version. The constants appear
in expressions for energy fluxes of waves and so I gave a derivation
in \cite{press1973} by looking at the asymptotic behavior of the fields
at large $r$, presumably the same derivation used by Starobinsky.
A little later I realized that the raising and lowering properties
hold in general, and not just at large $r$ \cite{teukolsky1974}.}

The identities described in this section are quite unexpected and
``miraculous.'' In spherical symmetry, we understand the existence
of raising and lowering operators on spin-weighted spherical harmonics
from the algebra of the rotation group. Here, by contrast,
equations following from certain laws of physics lead to the discovery
of identities that must be satisfied by a new class of special functions
of mathematical physics, the spin-weighted spheroidal harmonics and the
related radial functions. For a discussion of the identities from
the special-function point of view, see \cite{fiziev2009}.

\subsection{Superradiant scattering}
\label{sec:superradiant}

Section \ref{sec:penroseproc} described the Penrose process for particles.
There is also its wave analogue, \emph{superradiance} \cite{zeldovich1971}.
For wave modes proportional to $\exp{-i(\omega t-m\phi)}$,
a scattered wave can come off with more energy than the incident wave if
$m > 0$ and $0<\omega < m \Omega$. (The angular velocity
of the horizon $\Omega$ was defined in 
\eref{eq:angvel}.)
Essentially some negative energy flux is absorbed by the hole.

The existence of superradiance made it seem more likely that Kerr
black holes might be dynamically unstable.
Numerical calculations \cite{press1972,teukolsky1974} showed that
the maximum superradiant amplification of low-$m$ modes is bounded,
reaching 138\% for the $l=m=2$ gravitational mode. 
The stability question is taken up in the next subsection.

Superradiance was invoked to design the ``black-hole bomb'' \cite{press1972},
where a mirror around the black hole reflects radiation back toward
the hole multiple times. The amplitude grows exponentially until
the mirror explodes. Leaving aside the possibility that an advanced
civilization might be able to control the power flow through suitable
ports in the mirror, this scenario suggests that a scalar field with
mass would be unstable near a Kerr black hole. Negative energy could
radiate down the hole, but no energy would leave at infinity because
the solution itself decays exponentially fast. Such instability
has in fact been rigorously demonstrated \cite{rothman2014}.

\subsection{Stability of Kerr black holes}

\subsubsection{Mode stability}
As already mentioned, the primary motivation of the early work on Schwarzschild
perturbations was to prove that spherical black holes were stable objects,
and hence might actually exist in the physical universe.
After separating out the angular dependence, for modes
of the form
\begin{equation}
\psi=R(r)e^{-i\omega t}
\label{eq:modes}
\end{equation}
both the Regge-Wheeler and Zerilli equations can be written in the
form of Schrodinger-like equations:
\begin{equation}
\frac{\rmd^2 R}{\rmd r^{*2}} -V(r)R=-\omega^2 R.
\label{eq:zerilli}
\end{equation}
Here $r^*(r)$ is the so-called tortoise coordinate that maps the interval
$(r_+,\infty)$ to
the interval $(-\infty,\infty)$.
The potential $V$ goes to zero at both ends of the interval
and is real, positive,  and independent of $\omega$.
There are two standard arguments to show the stability of modes satisfying
\eref{eq:zerilli}.

First, \eref{eq:zerilli} is a linear, self-adjoint eigenvalue problem
for $\omega^2$. By self-adjointness, $\omega^2$ must be real, so any
instability must lie on the positive imaginary axis, $\omega=i\alpha$ say.
Thus the equation takes the form
\begin{equation}
\frac{\rmd^2 R}{\rmd r^{*2}} = \textrm{(positive-definite function)} R
\end{equation}
which manifestly has no solution that is regular at $r^* \to \pm\infty$.

A second way of proving stability is to note that the modes \eref{eq:modes}
satisfy the wave equation
\begin{equation}
\frac{\partial^2\psi}{\partial r^{*2}} -
\frac{\partial^2\psi}{\partial t^2} -V(r)\psi=0.
\label{eq:waveeqn}
\end{equation}
Associated with this equation is the energy
\begin{equation}
E=\frac{1}{2}\int_{-\infty}^\infty \rmd r^*\left(
\left|\frac{\partial\psi}{\partial t}\right|^2+
\left|\frac{\partial\psi}{\partial r^*}\right|^2
+V|\psi|^2\right).
\label{eq:waveenergy}
\end{equation}
Since the potential is nonnegative, so is the energy.
The boundary conditions for well-behaved solutions of \eref{eq:waveeqn}
are
\begin{equation}
\psi \to \cases{A e^{-i\omega(t-r^*)},& $r^*\to \infty$\\
B e^{-i\omega(t+r^*)},& $r^*\to -\infty$.}
\label{eq:bcs}
\end{equation}
Physically, these correspond to waves leaving the domain at both infinity
and the event horizon.
Evaluating
$\rmd E/\rmd t$, simplifying using the wave equation \eref{eq:waveeqn},
and integrating by parts gives
\begin{equation}
\frac{\rmd E}{\rmd t}= -|\omega|^2(|A|^2+|B|^2).
\label{eq:fluxes}
\end{equation}
Thus the energy cannot grow without bound, so the terms in the integrand
of \eref{eq:waveenergy} are bounded, proving stability.

For the Kerr metric, the effective potential associated with the radial
equation \eref{eq:radial} is not real nor independent of
$\omega$. Thus there does not seem to be a stability proof
analogous to the self-adjointness argument for Schwarzschild.
Turning to the second method of proving stability, one can write down
a conserved energy
based on \eref{eq:master}. However, the energy density
is not positive in
the ergosphere. Thus the total energy can be finite while the field
still grows exponentially in parts of spacetime.
Press and Teukolsky \cite{press1973} noted that
boundary conditions analogous to \eref{eq:bcs} imply that an instability
corresponds to the incoming wave amplitude from
infinity having a zero for $\omega$ in
the upper half-plane. Since Schwarzschild is stable, for $a=0$ all such
zeros must be in the lower half-plane.
Assuming that an instability could occur only
if a zero migrated smoothly from the lower half plane as $a$ increases from
zero, one needs to search only the real axis for various values of $a$,
which we did numerically for the lowest lying angular modes. Paper
\cite{teukolsky1974} found a conserved energy for the radial
equation \eref{eq:radial} from the Wronskian of two linearly independent
solutions and hence showed that an instability could occur only by a
zero migrating in the finite superradiant
frequency range between 0 and $m\Omega$. Hartle and Wilkins
\cite{hartle1974} proved that the assumption of smooth migration was
valid, but stability still rested on the numerical search results.
Finally, in 1989, Whiting \cite{whiting1989}
gave a rigorous proof that there could be no exponentially growing
modes in Kerr. In the proof, Whiting constructed a ``miraculous''
conserved quantity with a positive definite integrand.

For most physicists, this was the end of the story. Black holes were
unequivocally
established as stable objects firmly predicted by general relativity
and likely to be found in various astrophysical settings.
For the more mathematically inclined, the story was only just beginning.

\subsubsection{Linear and nonlinear stability}
While an unstable mode implies that the system is unstable, showing that all
the modes are stable does not necessarily imply that the system is stable.
One needs to be sure that any perturbation can be expressed as
a superposition of modes. This completeness is guaranteed for self-adjoint
problems like Schwarzschild, but requires further work in Kerr. More
important, superposing an infinite number of stable Fourier modes
does not guarantee that the result is stable. 
The standard method of proving stability in this case relies on
a conserved energy like \eref{eq:waveenergy}. But instead of relying on
modes like \eref{eq:bcs} to infer that the energy does not grow as
is done in \eref{eq:fluxes}, one has to find quantitative estimates of the
decay of the field in time to show boundedness of the energy.

Showing boundedness of perturbations is called proving linear
stability, which is all that is necessary for linear equations.
Boundedness of all solutions of a nonlinear system gives nonlinear stability.
Often, proving linear stability first makes it easy to prove
nonlinear stability.

The first full linear stability proof for Schwarzschild
was carried out by Kay and Wald \cite{kay1987},
but the Kerr case is still not completely done.
See  \cite{dafermos2008,dafermos2010,dafermos2014}
for status reports and references.

\subsection{Metric reconstruction}
In Schwarzschild, the Regge-Wheeler and Zerilli equations directly give
the odd- and even-parity perturbations for certain combinations of the metric.
Given solutions of these equations, the full metric can be reconstructed
from the remaining perturbation equations. In Kerr, by contrast,
the perturbation equation \eref{eq:master} describes either the $\Psi_4$
or $\Psi_0$ component of the Weyl tensor. Since this tensor consists
of second derivatives of the metric, reconstructing the corresponding
metric perturbation thus requires two integrations. Moreover, finding
the complete metric involves solving the remaining Newman-Penrose
equations. And finally, $\Psi_4$ and $\Psi_0$ have spin weight 2
and so give no information about $l=0$ and $l=1$ metric perturbations
corresponding to shifts in mass and angular momentum;
they must be separately reconstructed.

Chandrasekhar \cite{chandra1983} has carried out this reconstruction
in an amazing feat of analysis. The procedure is so complicated that
it does not seem to have been used, at least in its entirety, in
any application. Attention has instead focused on an alternative
strategy that uses the analogue of Hertz potentials. This idea was introduced
by Chrzanowski \cite{chrzanowski1975}  and then by a somewhat different
route by Cohen and Kegeles \cite{cohen1975,kegeles1979},
following their earlier work
on the electromagnetic case \cite{cohen1974}. 
Later work by Wald \cite{wald1978} and Stewart \cite{stewart1979}
further elucidated the method.
The Hertz potential $\Psi$ satisfies the
homogeneous master equation \eref{eq:master}, but
it is not the 
$\Psi_0$ or $\Psi_4$ arising from the metric perturbation it generates.
Instead,
the actual reconstruction procedure is as follows:
\begin{enumerate}
\item
Solve the master equation for $\psi$ with the physical boundary conditions
and source, where $\psi$ is one of the quantities $\Psi_0$ or $\Psi_4$.
\item
Solve an equation of the form $\mathcal{D}\Psi^*=\psi$, where
$\mathcal{D}$ is a certain
fourth-order partial differential operator and $*$ denotes complex
conjugation.
This equation can be solved by separation of variables
using spin-weighted spheroidal harmonics.
The Hertz potential $\Psi$ must also satisfy the
homogeneous master equation \eref{eq:master}.
This procedure was first explicitly carried out for Kerr by Ori \cite{ori2003}.
\item
\label{item:chrz}
Compute the metric by applying a certain second-order partial differential
operator to $\Psi$. 
(This is the formula originally given by Chrzanowski, Cohen, and Kegeles
\cite{chrzanowski1975,cohen1975,kegeles1979}.) 
\item
Complete the solution by finding and adding the $l=0$ and $l=1$ pieces
to the metric. This part of the procedure does not seem to have
a general formulation, and so far is carried out case-by-case.
\end{enumerate}
The resulting metric is given in the so-called ``radiation gauge.''
When sources are present, the gauge leads to singularities even
in vacuum regions, and a great deal of effort has been expended in
dealing with these complications.

A nice pedagogical example of metric reconstruction is given in
\cite{sano2014}. The principal application of reconstruction
has been to the self-force problem for small bodies interacting with Kerr
black holes. See \cite{poisson2011} for a review and references.

\subsection{Black hole quasi-normal modes}
\label{sec:qnm}
If a black hole is perturbed in some way, the radiation produced generally
has three components. At early times,
there is prompt emission that reflects the details
of the source and whose Fourier components basically propagate out
along null rays. At very late times, there is ``tail'' emission produced
by backscatter of the prompt radiation
off the curvature of spacetime. The amplitude of tail emission decays
as a power law in time, as shown originally
by Price \cite{price1972a,price1972b}
for the Schwarzschild case. In between these two components the
radiation is dominated by the
black hole \emph{ringdown}. This can be thought of as radiation
produced by the dynamical excitation of the black hole spacetime, which
then emits radiation much like a struck bell rings down by emitting
certain characteristic tones. Like a real bell, a black hole system is
dissipative since the waves carry off energy. Accordingly, the
modes have an imaginary component in the frequency that represents
the damping rate.

Ringdown was noticed very early in studies of perturbations of
Schwarzschild black holes \cite{vishveshwara1970b,delacruz1970}.
In 1971 Press \cite{press1971}
introduced the viewpoint that ringdown waves
should be thought of as the free oscillations of the black hole and
called them quasi-normal modes.
Since then many applications of these modes have been found
and a huge literature has developed. A comprehensive review is
provided in \cite{berti2009}, which references a number of other
more specialized reviews.
Here I will describe just a few important applications.

First, quasi-normal modes are a ubiquitous feature of gravitational radiation
from astrophysical black holes, and searches by detectors such as LIGO and
VIRGO try to include their effects.
In this case we expect that only the low-lying least
damped modes are important. We may be lucky enough to measure
the frequencies and decay times for two modes from, for example, the
inspiral and merger of two black holes. Since the frequency and decay time
depend only on the mass and spin of the final black hole, with four
measured quantities we would
have a very clean test of general relativity, namely that the final state
is indeed described by the Kerr metric.

A prescient theoretical
application of quasi-normal modes was the paper by York
in 1983 \cite{york1983}, who showed how the statistical mechanics
of the modes was related to the black hole entropy.
Further interest was sparked by Hod \cite{hod1998} in 1998, who used
asymptotic values for the highly damped modes to conjecture the spacing of
the area spectrum for quantized black holes. Soon after this,
Horowitz and Hubeny \cite{horowitz2000} showed
how quasi-normal modes could be used in the AdS/CFT correspondence:
modes for black holes in AdS space are related to relaxation
times in the dual conformal field theory. Since then there has been
an explosion of work exploiting quasi-normal modes to elucidate
the AdS/CFT correspondence and quantum gravity (see
\cite{berti2009,konoplya2011} for reviews and references).

\subsection{Analytic solutions}
\label{sec:analytic}

Analytic treatment of the perturbation equations \eref{eq:angular} and
\eref{eq:radial} was initiated in an important paper by Leaver \cite{leaver1985}
on Kerr quasi-normal modes.
Leaver noted that the equations were similar to those solved by
Jaff\'e in 1934 \cite{jaffe1934} to determine the electronic spectrum
of the hydrogen molecule ion.
In particular,
the radial equation \eref{eq:radial} has two regular singular points,
at the horizon radii $r_+$ and $r_-$, the roots of $\Delta=0$. It also
has an irregular singular point at infinity. Leaver
constructed solutions as expansions about the horizon,
\begin{equation}
R(r) \sim \sum_{n=0}^\infty a_n z^n, \qquad
z\equiv\frac{r-r_+}{r-r_-}
\label{eq:leaver}
\end{equation}
where the $\sim$ symbol means I have left out a known prefactor function
of $r$ for simplicity. The prefactor includes the boundary conditions
of ingoing waves at $r=r_+$ and outgoing waves at infinity.
Equation \eref{eq:leaver} will be a valid solution all the
way to $r=\infty$ only if the series
is convergent for any $z\in[0,1]$.
Substituting the series into the radial equation
leads to a three-term recurrence relation for the coefficients:
\begin{eqnarray}
\label{eq:recur0}
\alpha_0 a_1 +\beta_0 a_0 = 0\\
\alpha_n a_{n+1} + \beta_n a_n + \gamma_n a_{n-1} = 0, \qquad n=1,2, \ldots
\label{eq:recur}
\end{eqnarray}
Here the coefficients $\alpha_n$, $\beta_n$, and $\gamma_n$ depend
only on the parameters in the differential equation.
Investigating the behavior of $a_n$ in \eref{eq:recur} for large
$n$ shows that of the two linearly independent solutions, only
the one for which $a_n$ decreases with $n$ will allow the series
to converge. The standard
theory of three-term recurrence relations (cf., e.g., \cite{press2007}
or \cite{gautschi1967}) says that the corresponding solution
of \eref{eq:recur} is the so-called minimal solution.
This solution can be found from the continued fraction that results
from rewriting \eref{eq:recur} for the ratio $a_{n+1}/a_n$ and iterating:
\begin{equation}
\frac{a_{n+1}}{a_n}=\frac{-\gamma_{n+1}}{\beta_{n+1}-{}}\,
\frac{\alpha_{n+1}\gamma_{n+2}}{\beta_{n+2}-{}}\,
\frac{\alpha_{n+2}\gamma_{n+3}}{\beta_{n+3}-{}}
\dots
\label{eq:cf}
\end{equation}
Evaluating \eref{eq:cf} for $n=0$ and equating this to $a_1/a_0$ from
\eref{eq:recur0} gives an implicit equation for the characteristic
frequency of the normal mode that can be solved numerically.
For other values of $\omega$, the
equation will not be satisfied.

In the Schwarzschild case, the eigenvalue of the angular equation is
known analytically, $A(a\omega=0)=l(l+1)-s(s+1)$,
and is independent of $\omega$.
In the Kerr case, the angular equation has to be solved simultaneously
with the radial equation to determine $A$. As a function of
$u=\cos\theta$, the angular equation also
has two regular singular points and one irregular singular point.
Thus it can be handled in exactly the same way as the radial equation,
this time as a series in $z=1+\cos\theta$. The three-term recurrence
relation leads to a continued fraction for the minimal solution and
hence an implicit equation for $A$. So Kerr modes can be found
by solving the two implicit equations simultaneously for $A$ and $\omega$.

Leaver \cite{leaver1986} also found several other representations of
solutions of the perturbation equations. The most useful 
is an expansion as a series of Coulomb wave functions. The coefficients
in the expansion once again lead to a three-term recurrence relation,
but this time the expansion index $n$ ranges from $-\infty$ to $\infty$.
These solutions are convergent for $r_+ < r \leq \infty$.

The power series expansions \eref{eq:leaver} valid near the horizon
converge quite slowly. Mano, Suzuki, and Takasugi \cite{mano1996}
instead found expansions in hypergeometric functions that are
valid for $r_+ \leq r < \infty$. The expansion coefficients have the \emph{same}
three-term recurrence relation as the Coulomb wave function expansion, which
is valid from infinity inward. They made some improvements in \cite{mano1997},
and a summary can be found in \S4 of \cite{sasaki2003}. A key element of
the analysis was a set of connection formulas that allow one to match
the expansion valid near the horizon to the expansion valid at infinity,
and thus impose appropriate boundary conditions.
A number of applications are given in \cite{sasaki2003}. A recent application
I cannot help mentioning is that of Fujita \cite{fujita2012}, who
computed the analytic post-Newtonian expansion of the gravitational waveform
for a particle in circular orbit about a Schwarzschild black hole to
22nd order! This kind of analysis is useful not just for waveforms of
extreme mass-ratio inspiral binaries, but also for studying
the range of validity of PN expansions, which is not well understood at all.

Instead of solving the angular equation as
a power series in $1+\cos\theta$, one can make an expansion in Jacobi
polynomials, again leading to a three-term recurrence relation. This
method is due to Fackerell and Crossman \cite{fackerell1977}, with some
improvements in \cite{fujita2004}.

Ordinary differential equations like the radial and angular perturbation
equations are examples of the confluent Heun equation. There has been
growing interest in analyzing the equations from this point of view (see e.g.,
\cite{fiziev2010}). A result that emerges from this analysis is that
the radial and angular equations are essentially ``the same,'' related
by (possibly complex) coordinate transformations. This is the reason
that the identities of \S\ref{subsec:teukstar} exist for both
the angular and radial functions: If they exist for one variable,
they must exist for the other.

\section{``Explanation'' of the miracles of Kerr: the Killing-Yano tensor}
Everyone who spends time working with the Kerr metric comes away
convinced that that there is something magical about it.
There is first the very fact of its existence. 
We may not be surprised that the special features of Type D metrics
allow us to find analytic solutions, but there is no a priori reason
why the most general rotating black hole solution should be a member of
this class. There are also small surprises, like the unexpectedly
simple formula for the angular velocity of a particle in
an equatorial circular orbit,
\begin{equation}
\Omega_{\rm circular}=\pm \frac{M^{1/2}}{r^{3/2}\pm aM^{1/2}}.
\end{equation}
Here the upper and lower signs refer to prograde and retrograde orbits.
Only a simple modification of Kepler's Third Law to incorporate rotation
is required.

But the real feeling that something is special begins with encountering
the unexpected separability of the Hamilton-Jacobi and scalar wave
equations. The additional integral of motion provided by Carter's constant
makes the geodesic equations completely integrable. Then an even bigger
surprise, the decoupling and separability of the equations for
electromagnetic, neutrino, and gravitational perturbations.
And among the various functions encountered in the solution of these
perturbation equations, a panoply of identities and relationships
required by the Maxwell or Einstein equations and in fact
satisfied by the solutions as new ``special functions''
of mathematical physics.
No wonder Chandrasekhar \cite{chandra1983} was led to refer to the
``\dots many properties which have endowed the Kerr metric
with an aura of the miraculous.''

The existence of $t$ and $\phi$ constants of the motion or separability
follows directly from the stationarity and axisymmetry of the metric,
or equivalently from the existence of the two Killing vectors
$\partial_t$ and $\partial_\phi$. But there is no group theory
to account for
the separability in $r$ and $\theta$. The separation is characterized by
Carter's constant, which is quadratic in the
particle momenta and generalizes the total angular momentum of Schwarzschild.
In 1970, Walker and Penrose \cite{walker1970}
showed that its existence follows from
the existence of a symmetric Killing tensor for Kerr that satisfies
a generalization of Killing's equation
\begin{equation}
K_{(ab;c)}=0.
\end{equation}
The Killing tensor, in turn, is the ``square'' of the
antisymmetric Killing-Yano tensor \cite{penrose1973,floyd1973},
which satisfies
\begin{equation}
K_{ab}=f_{ac}f_b{}^c,\qquad f_{a(b;c)}=0.
\end{equation}
A metric that admits a Killing-Yano tensor must be Type D \cite{collinson1974},
but not all Type D metrics actually have one \cite{demianski1980}.

Carter \cite{carter1977} showed that in a vacuum geometry,
a Killing vector or Killing tensor
can be used to construct operators that commute with the scalar Laplacian
and hence imply its separability.
Carter and McLenaghan \cite{carter1979}
next constructed an operator from the Killing-Yano
tensor that commutes with the operator in the Dirac equation,
explaining its separability.
However, the separability of the higher spin equations is still
mysterious. The separation constant itself has been characterized
as the eigenvalue of an operator constructed from the Killing-Yano tensor
\cite{kalnins1989,kalnins1996}, but commutation relations or
similar criteria that would
imply separability have not been found.
The situation is summarized in \cite{kalnins1999}.

Even if one could show that the Killing-Yano tensor implies all the
separability properties of Kerr, it is not clear in what sense this
would be an explanation rather than a restatement.
Symmetry or group theory is an explanation of separability
because they are general and fundamental properties, but the existence of
the Killing-Yano tensor seems to be a miracle on the same level
as separability.

Another way of seeing why the Killing-Yano tensor explanation is
inadequate is to consider the remarkable identities discussed in
\S\ref{subsec:teukstar}. In spherical symmetry, the angular identities
are related to the raising and lowering operators for spin-weighted
spherical harmonics that follow from group theory. The radial identities
then follow by mapping the angular equation to the radial equation,
as mentioned at the end of \S\ref{sec:analytic}.
In Kerr, however,
these identities emerge as unexpected delights, with no known
deep reason for their existence.

Many of the above results on ``hidden symmetries''
for the Kerr metric can be extended to more
general spacetimes or to higher dimensions (see \cite{frolov2008} for
a discussion and references).

\section{The solution enters astrophysics}
\label{sec:astro}

The first Texas symposium in 1963 was spurred by the discovery of quasars
and the belief that relativity might have something to do with explaining
them, but it was not until a few years later that the discipline of 
black hole astrophysics was really born. The discovery of pulsars
in 1967 \cite{hewish1968} and their quick identification as rotating
neutron stars \cite{gold1968}
introduced the first highly relativistic objects into astrophysics.
The launch of X-ray satellites dedicated to astronomy starting with
Uhuru in late 1970 led to a spate of discoveries of compact objects
accreting matter from companion stars. Among these X-ray sources
was Cyg X-1, the first reliable stellar-mass black hole candidate.
Soon after this, strong observational evidence began to emerge that
quasars and other powerful radio sources at centers of galaxies
are powered by supermassive black
holes. By now relativistic astrophysics is a vast subject, and here
I will focus only on the central role of the Kerr metric and on
attempts to verify that Kerr black holes actually exist with all the
properties predicted by general relativity.

\subsection{Observations of black holes}

\subsubsection{Masses of stellar-mass black holes}
For a stellar-mass compact object,
the main argument used to claim that it is a black hole is to determine
its mass reliably. If this mass is greater than the maximum mass of
a neutron star, then the object must be a black hole.

The maximum mass of a neutron star is uncertain because of our
lack of understanding of the properties of nuclear matter at high densities.
However, it is generally believed to be less than $2.5\,M_\odot$,
with an absolute upper bound around $3\,M_\odot$ (for a recent review
and references, see \cite{chamel2013}). The most massive reliable
observed neutron
star mass is $2.01\pm 0.04\,M_\odot$ \cite{antoniadis2013}.

All the accurate black hole
mass determinations come from X-ray binary systems.
Gas flows from the companion star onto a compact object through
an accretion disk. (Here compact object means a white dwarf, neutron
star, or black hole. A normal star is ruled out by the small size of the
binary orbit.) The hot disk emits X-rays. The mass of the compact
object is measured
by a venerable technique of classical astronomy, measuring the
radial velocity curve of the companion by the Doppler shift
of its spectral lines. From the velocity curve and Kepler's Third Law
one gets the mass function
\begin{equation}
f(M)=\frac{M\sin^3i}{(1+q)^2}.
\label{eq:massfunction}
\end{equation}
Here $i$ is the orbital inclination of the binary to the line of sight
and $q$ is the ratio of companion mass to black hole mass $M$. Determining
$M$ requires further observational and theoretical inputs to pin down
$i$ and $q$. This has been done for about two dozen sources where $M$
is accurately measured to be greater than $3M_\odot$
\cite{casares2013,ozel2010}, the
maximum neutron star mass.

Note that one can rewrite \eref{eq:massfunction} as
\begin{equation}
M=\frac{(1+q)^2}{\sin^3i}f.
\end{equation}
Since $q\geq 0$ and $\sin i \leq 1$, the \emph{minimum} mass is given by
the value of the mass function $f$, an extremely robust observable.
About ten of the known compact X-ray sources have values of $f$ in the
range 3 -- $8\,M_\odot$. There is no wiggle room to argue about
uncertainties in astrophysical modeling; these sources must be black holes.

\subsubsection{Spins of stellar-mass black holes}

Black hole spin measurements have nowhere near the reliability of
the precise mass measurements possible using orbital dynamics.
Determining the spin
requires difficult observations from matter very close to the event horizon,
where the metric is most affected by the value of $a$.

For stellar-mass black holes, the continuum fitting
method relies on models of the
accretion disk. The gravity produced by the disk is negligible compared
with that of the black hole, so the model can be constructed in a fixed
Kerr background metric. In the thin disk model, the inner edge of
the disk occurs at the innermost stable circular orbit
$r_{\rm ISCO}$, a simple function of $a$
\cite{bardeen1972}. Fitting the continuum spectrum of
the disk to the thin-disk model thus gives an estimate of $a$. About ten
black hole spins have been estimated in this way, with
values ranging up to near extremal. For more details, see for example
\cite{mcclintock2013}.

A second method for determining the black hole spin
is relativistic X-ray reflection spectroscopy, also called
broad iron line spectroscopy. A number of compact X-ray sources show
a hard component in the spectrum that comes from a ``corona'' just
above and below the accretion disk. The irradiated disk produces
a reflection signature in the X-ray spectrum, most prominently
an iron-K$\alpha$ emission line. The line is Doppler and gravitationally
redshifted by different amounts at different radii in the disk, and
fitting a model to the radiation gives an estimate of $a$.
More than a dozen spins have been measured in this way. For
more details, see \cite{reynolds2013a}. In a few cases
measurements have also been made on the same sources
by the continuum fitting method, and
the results are roughly consistent. However, both methods are still
subject to quite large uncertainties.

\subsubsection{Masses of supermassive black hole}

Estimating the mass of a supermassive black hole  can be done in several
ways. For example, the radiation is produced by accretion, with the luminosity
limited by the Eddington limit at which the outward radiation pressure
is balanced by the input pull of gravity,
\begin{equation}
L_{\rm E}=\frac{4\pi cGMm_p}{\sigma_{\rm T}} \sim 10^{38}\frac{M}{M_\odot}
\,{\rm erg}\,{\rm s}^{-1}.
\end{equation}
Here $m_p$ is the proton mass and $\sigma_{\rm T}$ the Thomson cross
section for electron scattering. Bright quasars have luminosities
$L \sim 10^{46}\,{\rm erg}\,{\rm s}^{-1}$, so $M \gtrsim 10^8M_\odot$.

For such an object to be a black hole, its size has to be sufficiently
small. Quasars show variability on a timescale of days. Setting this timescale
to the light crossing time implies that the object is smaller than
a few light days, or ${}\sim 100 M$. Even tighter limits for some
sources come from Doppler measurements of X-ray lines, showing that
matter is orbiting at a substantial fraction of $c$ \cite{fabian2000}.

The black hole at the center of our Galaxy, Sgr $\rm A^*$, has
a very well-determined mass.
(For a review
of the arguments that this radio source is the black hole, see, e.g.,
\cite{narayan2013}.) The most precise value
to date is $(4.26 \pm 0.14)\times 10^6\,M_\odot$ \cite{chatzopoulos2014},
determined from a combined analysis of individual stellar orbits
around the black hole and the dynamics of the nuclear star cluster.

\subsubsection{Spins of supermassive black holes}
Spins of supermassive black holes are typically determined
using the X-ray reflection spectroscopy method, which requires
the presence of a bright accretion disk. (The continuum
fitting method is difficult to apply in the supermassive case.)
About two dozen spins have been estimated \cite{reynolds2013b,reynolds2013a},
many suggesting that rapidly spinning massive
Kerr black holes do actually exist.

Since Sgr $\rm A^*$ does not have a bright X-ray emitting accretion disk,
the prospects of distinguishing between Schwarzschild and Kerr
for its metric are not good at present.

\subsection{Energy extraction from astrophysical black holes}
\label{sec:bz}
What exactly is the mechanism by which supermassive black holes at
the centers of galaxies power such energetic phenomena as quasars, AGNs,
and relativistic jets? And what about gamma-ray bursts and relativistic
outflows in galactic X-ray binaries? It is widely believed that the
Blandford-Znajek process \cite{blandford1977} is one of the most promising
mechanisms in many of these cases.

The Blandford-Znajek process is related to the Penrose process, in
that rotational energy is extracted from the black hole because
negative energy can flow down the hole if there is an ergosphere.
However, it is a purely electromagnetic process---plasma is present
to support the fields,
but its inertia and energy are negligible.
Moreover, the presence of plasma allows the process to work even
with stationary fields.

In the 1990's, there was some controversy about whether the Blandford-Znajek
process was actually theoretically viable, but that has been resolved
and the process itself is now well understood
(see, e.g., \cite{komissarov2009,gralla2014,toma2014,koide2014,lasota2014}).
However, complete and convincing models of the accretion, jet formation, and
energy emission are still beyond the reach of
current large-scale numerical simulations.

\subsection{Are astrophysical black hole candidates really black holes?}
The astronomical observations described above are all of the form
``there are regions of the Universe with a lot of mass in a small
volume.'' \emph{If} we assume general relativity is correct, then
these must be black holes. But there are still no convincing observations
that tell us directly that these objects have the bizarre properties predicted
by relativity, such as event horizons or an exterior geometry described
by the Kerr metric. As long as we depend only on electromagnetic observations,
we will be at the mercy of the uncertainties in the theoretical models
of the radiation.

This situation is about to change. Within a few years, gravitational wave
detectors such as LIGO and VIRGO will reach a sensitivity at which
they are expected to detect waves from the inspiral and merger of
binary systems containing black holes and neutron stars. In particular,
waves from a black hole-black hole merger will probe for the first time
the strong-field regime as the black holes merge. Comparison with
numerical solutions of the full Einstein equations will provide
powerful tests of general relativity in this regime. And measurement of the
ringdown waves will show directly the settling down of the final
state to a Kerr black hole. Or maybe to something unexpected \dots\ .



\ack
I thank Leo Stein, Stanley Deser and Ted Jacobson for helpful discussions.
This work was supported in part by
NSF Grants PHY-1306125 and AST-1333129 at Cornell University, and
by a grant from the Sherman Fairchild Foundation.




\providecommand{\newblock}{}


\end{document}